\documentstyle[preprint,aps]{revtex}
\begin{document}
\draft
\title{\bf SCHILD'S NULL STRINGS IN FLAT AND CURVED BACKGROUNDS}
\author{Sayan Kar \thanks{Electronic Address :
sayan@iopb.ernet.in}}
\address{Institute of Physics\\
Sachivalaya Marg, Bhubaneswar--751005, INDIA}
\maketitle
\begin{abstract}
Schild's null (tensionless) strings are discussed in certain flat and
curved backgrounds. We find closed, stationary, null strings
as natural configurations existing on the horizons of
spacetimes which possess such null hypersurfaces. Examples of these are
obtained in Schwarzschild and Rindler spacetimes. A dynamic
null string is also identified in Rindler spacetime.
Furthermore, a general prescription (with explicit examples)
is outlined by means of which
null string configurations can be obtained in a large class of
cosmological backgrounds .
\end{abstract}
\vspace{.3in}

\newpage

In a posthumously published paper, about eighteen years ago,
Schild {\cite{as:prd77}} first introduced the notion of a null
or tensionless string (see also Karlhede and Lindstrom{\cite{kl:cqg86}})
. By replacing
the Nambu--Goto Lagrangian with its square he wrote down the
action and equations of motion for the null string following the
standard treatment for null geodesics. The null string was
quantized by Lizzi et. al. {\cite{lsrs:plb86}}. Surprisingly
, they found that consistent quantization did not imply a critical
dimension. Subsequently, it was realized that the absence of a
critical dimension was essentially an artefact of the ordering
of operators in the quantum theory. Whereas Lizzi et. al.
preferred Weyl ordering (the absence of oscillator modes being
the rationale) others {\cite{bnra:plb89}},{\cite{grra:npb90}}
used normal ordering and obtained $D=26$ as the critical
dimension. Simultaneously, a Hamiltonian analysis was also
carried out {\cite{aaz:sjnp88}} and supersymmetric
generalisations were discussed {\cite{grra:npb90}}. The
latest, carefully done Hamiltonian BRST analysis due to
Gustaffson et. al. {\cite{glssu:npb95}} which employs the
use of a smeared delta function in the canonical commutation relations
 concludes that the critical
dimension of the bosonic null string is $D=2$. A review of the
literature on null
p--branes and super p--branes upto 1993 can be found in Bandos and
Zheltukhin {\cite{bz:fdp93}}.

However, null strings have almost never been discussed in a
curved Lorentzian background. The only reference to a curved
background can be found in the concluding portion of Schild's
paper. More recently, however it has been shown {\cite{ar:grqc95}}
that an energy momentum tensor describing a fluid of null strings
can act as a source for metrics representing
Friedman-- Robertson--Walker universes in both its matter
and radiation dominated epochs. In this article, we first set up the
equations of motion of a null string in a curved background and then
obtain exact
string configurations in a variety of backgrounds which include
the Rindler, Schwarzschild and some cosmological spacetimes.

Let us begin with the curved background action written down by
Schild {\cite{as:prd77}}. This is given as:

\begin{equation}
S = \int \Sigma^{2} d\sigma d\tau
\end{equation}

where $\Sigma^{2} =
g_{\mu\alpha}g_{\nu\beta}{\Sigma}^{\mu\nu}{\Sigma}^
{\alpha\beta}$ and

\begin{equation}
{\Sigma}^{\mu\nu} = \frac{\partial x^{\mu}}{\partial \tau}
\frac{\partial x^{\nu}}{\partial \sigma} - \frac{\partial x^{\mu}}{\partial
\sigma}\frac{\partial x^{\nu}}{\partial \tau}
\end{equation}

Here $x^{\mu}(\tau,\sigma)$ is the embedding function for the
null world--sheet in a general background spacetime.

Note that $\Sigma^{2}$ is essentially the determinant of the
metric induced on the worldsheet by the background spacetime.
If the worldsheet is null then there must exist one null tangent
vector. This leads to the fact that the determinant of the
induced metric is identically zero.

The field equations and constraints that arise out of variations
of the action in (1) are given as:

\begin{equation}
 {\ddot x}^{\mu} +
{\Gamma}^{\mu}_{\rho\lambda} {\dot x}^{\rho}{\dot x}^{\lambda} = 0
\end{equation}

\begin{equation}
g_{\mu\nu}{\dot x}^{\mu}{\dot x}^{\nu} = 0 \qquad ; \qquad
g_{\mu\nu}{\dot x}^{\mu}{x^{\prime}}^{\nu} = 0
\end{equation}

where the overdots and primes denote differentiation with
respect to $\tau$ and $\sigma$ respectively.

The constraint equations are not invariant for arbitrary
reparametrizations of $\sigma$ and $\tau$. In fact the allowed
reparametrizations are $\tau_{1} = f(\sigma, \tau)$ and
$\sigma_{1} = g(\sigma)$. Thus the $\sigma\sigma$ element of the
induced worldsheet metric has to be independent of $\tau$. Then
only one has a reparametrization invariant theory where the
class of reparametrizations allowed are restricted in comparison
to the usual timelike bosonic string{\cite{bal:npb86}}.
Infact one can show that this restricted set of
transformations which leave the degenerate character of the
metric invariant form the Caroll group introduced by Levy
Leblond{\cite{ll:aihp65}} many years ago.

The field equations are essentially geodesic equations. The
constraint ($g_{\mu\nu}{\dot x}^{\mu}{\dot x}^{\nu} = 0$)
 implies that we should look for only null
geodesics. Thus, knowing the null geodesics in a background
spacetime would naturally lead to null string configurations
provided all the constraints are satisfied.  This is reminiscent
of the bosonic case where the stationary string equations can
also be transformed into geodesic equations in an {\em
unphysical} Riemannian space {\cite{sanch:conf}}
. It is also worth noting that the
equations of motion of a null string are not that of an extremal
surface in a background space of higher dimensions, as it is in
the case of the timelike bosonic string.

We now move on towards solving these equations of motion and
constraints in specific background spacetimes in order to obtain
specific string configurations.

{\em (a) Minkowski Spacetime}

To begin, let us deal with the almost trivial case of Minkowski
spacetime as the background. The null string equations of motion
and constraints yield a solution of the form :

\begin{equation}
x^{\mu} = a^{\mu}\tau + b^{\mu}(\sigma)
\end{equation}

with the constraints--

\begin{equation}
a^{\mu}a_{\mu} = 0 \hspace{.3in} and  \hspace{.3in} a^{\mu}b_{\mu}^{\prime}
(\sigma) = 0
\end{equation}

A simple choice for $a^{\mu}$ and $b^{\mu}$ could be

\begin{equation}
a^{\mu} \equiv (1, 1, 0, 0) \hspace{.3in} b^{\mu} \equiv (0, 0,
b^{2}(\sigma), b^{3}(\sigma))
\end{equation}

The string here is located on the lightcone of Minkowski spacetime
($x = t$). It can be closed or open depending on the choice of
$b^{2}(\sigma)$ and $b^{3}(\sigma)$.

{\em (b) Rindler Spacetime}

The metric in Rindler spacetime is given as:
\begin{equation}
ds^{2} = -a^{2}x^{2}dt^{2} + dx^{2} + dy^{2} + dz^{2}
\end{equation}

The null string equations of motion and constraints in this
background turn out to be :

\begin{equation}
{\ddot t} + 2\frac{\dot x}{x} {\dot t} = 0
\end{equation}

\begin{equation}
{\ddot x} + 2 a^{2}x{\dot t}^{2} = 0
\end{equation}

\begin{equation}
{\ddot y} = 0 \qquad ; \qquad {\ddot z} = 0
\end{equation}

\begin{equation}
-a^{2}x^{2}{\dot t}^{2} + {\dot x}^{2} + {\dot y}^{2} + {\dot
z}^{2} = 0
\end{equation}

Let us choose a generic string configuration as follows:

\begin{equation}
t = t(\tau) \quad ; \quad x = x(\tau) \quad ; \quad y =
y(\sigma) \quad ; \quad z = z(\sigma)
\end{equation}

A solution of the equations of motion and constraints is:

\begin{equation}
t = \frac{1}{2a} \ln \tau \quad ; \quad x = A_{1}
{\tau}^{\frac{1}{2}}
\quad ; \quad y = g_{1}(\sigma) \quad ; \quad z = g_{2}(\sigma)
\end{equation}

where $A_{1}$ can be any constant and $g_{1}(\sigma)$ and
$g_{2}(\sigma)$ are any two functions of $\sigma$. By
choosing the functions $g_{1}(\sigma)$ and $g_{2}(\sigma)$
suitably we can have different string configurations.
Moreover, as $\tau \rightarrow 0$ ($t\rightarrow -\infty$)
$x\rightarrow 0$ which is the event horizon of Rindler
spacetime. The domain $0\leq\tau\leq\infty$ encompasses
the whole of $-\infty\leq t\leq \infty$. The configuration
starts out at the horizon at $t\rightarrow-\infty$ and ends at
 $t\rightarrow\infty$
where $x\rightarrow \infty$
Notice also that the string equations of motion and constraints
have a very simple stationary string solution. This is given as:

\begin{equation}
t = \tau \quad ; \quad x = 0 \quad ; \quad y = f_{1}(\sigma)
\quad ; \quad
z = f_{2}(\sigma)
\end{equation}

This solution exists exclusively at the horizon of the Rindler
spacetime for all time $t$.  However this is quite expected because
the horizon
being a one way membrane is a null hypersurface of the
spacetime.

{\em (c) Schwarzschild Spacetime}

We now write down the string equations of motion and constraints
for a general static, spherically symmetric metric given by :

\begin{equation}
ds^{2} = -e^{2\Phi (r)} dt^{2} + \frac{dr^{2}}{1 - \frac{b(r)}{r}} +
r^{2}\left ( d\theta^{2} + \sin^{2} \theta d\phi^{2}\right )
\end{equation}

The null string equations of motion are :

\begin{eqnarray}
{\ddot t} + 2{\tilde \Phi}{\dot r}{\dot t} = 0  \\
{\ddot r} + \frac{{\tilde b} r - b}{2r(r-b)}{\dot r}^{2} -
(r-b) \left ( {\dot \theta}^{2} + \sin^{2}\theta {\dot \phi}^{2} \right ) = 0
 \\
{\ddot \theta} + \frac{2}{r} {\dot \theta}{\dot r} - \sin \theta \cos \theta
{\dot \phi}^{2} = 0 \\
{\ddot \phi} + \frac{2}{r} {\dot \phi}{\dot r} + 2 cot \theta {\dot \phi}
{\dot \theta} = 0
\end{eqnarray}

The constraint $g_{\mu\nu}{\dot x}^{\mu}{\dot x}^{\nu} = 0$
 reduces to the equation --

\begin{equation}
-e^{2\Phi}{\dot t}^{2} + \frac{{\dot r}^{2}}{1-\frac{b(r)}{r}} +
r^{2} {\dot \theta}^{2} + r^{2}\sin^{2}\theta {\dot \phi}^{2} = 0
\end{equation}

The other constraint is trivially satisfied because of the diagonal nature
of the background metric.
In addition, the quantity $g_{\mu\nu}x^{\mu\prime}x^{\nu\prime}$
must be independent of $\tau$.

It is easy to see that a closed stationary string of the form

\begin{equation}
t = \tau \quad ; \quad r = C \quad ; \quad \theta = \frac{\pi}{2}
\quad ; \quad \phi = C_{0} \sigma
\end{equation}

can exist in {\em any} such geometry which posseses a horizon
(i. e. $g_{00} = 0$
somewhere ) irrespective of the functional forms of $b(r)$ and $\Phi(r)$
 . In fact, the closed, stationary string exists exactly at the
horizon which is
by definition a null hypersurface. Thus, Lorentzian wormholes
cannot have a closed stationary null string anywhere. On the other hand,
they do support a closed, stationary timelike string at their throat as
has been pointed out in a recent paper by this author {\cite{sk:prd95}}.

{\em (d) Cosmological Spacetimes}

We now move on towards obtaining null strings in cosmological
models. The background metric is assumed as :

\begin{equation}
ds^{2} = -dt^{2} + R^{2}(t) \left [ \frac{dr^{2}}{1-kr^{2}} + r^{2} \right .
\left . (d\theta^{2} + \sin^{2}\theta d\phi^{2} ) \right ]
\end{equation}

where $ k = -1, 0, 1$ refer to hyperbolic, flat and three--sphere($S^{3}$)
spacelike sections.

The string equations of motion in this background turn out to be :

\begin{equation}
{\ddot t} + R\tilde R \frac{{\dot r}^{2}}{1-kr^{2}} + R\tilde R r^{2}
\left ( {\dot \theta}^{2} + \sin^{2}\theta{\dot \phi}^{2}\right ) = 0
\end{equation}

\begin{equation}
{\ddot r} + 2\frac{\tilde R}{R} {\dot r}{\dot t} + \frac{{\dot r}^{2}kr}
{1-kr^{2}} - (1-kr^{2})r \left ( {\dot \theta}^{2} + \sin^{2}\theta
{\dot \phi}^{2}\right )
= 0
\end{equation}

\begin{equation}
{\ddot \theta} + \frac{2}{r}{\dot r}{\dot \theta} + 2\frac{\tilde R}{R}
{\dot t}{\dot \theta} - \sin \theta \cos \theta {\dot \phi}^{2} = 0
\end{equation}

\begin{equation}
\dot {\overline {2 R^{2}r^{2}{\sin^{2}\theta}{\dot \phi}}} = 0
\end{equation}

and the constraint equation is :

\begin{equation}
-{\dot t}^{2} + R^{2}\left [ \frac{{\dot r}^{2}}{1-kr^{2}} + r^{2}
{\dot \theta}^{2} + r^{2}\sin ^{2}{\theta} {\dot \phi}^{2} \right ]
= 0
\end{equation}

Our ansatz for a string solution would be :

\begin{equation}
t = t(\tau) \quad ; \quad r = r(\tau) \quad ; \quad  \theta = \frac{\pi}{2}
\quad ; \quad \phi = C_{0} \sigma
\end{equation}

This represents a closed string which is dynamic (non--stationary).
The condition on $g_{\mu\nu} x^{\mu\prime}x^{\nu\prime}$ (i. e its being
independent of $\tau$) constrains the
choice of $r(\tau)$ and $R(t)$. We must have

\begin{equation}
r(\tau) R(t) = \frac{1}{C}
\end{equation}

where $C$ is some constant. Therefore, the radius $r(\tau)$ of the string
is inversely related to the scale factor governing the evolution of the
spacelike sections of the cosmological model. Note that by virtue of this
choice the degenerate induced metric on the null world--sheet is the same
for all forms of $R(t)$ or $r(\tau)$.

With the ansatz for a string configuration and subsequently
that of $r(\tau)$ we end up with the following two equations which
we have to solve in order to obtain $R(t)$ and $t(\tau)$.

\begin{equation}
\tilde R = \pm \sqrt {C^{2}R^{2} - k}
\end{equation}

\begin{equation}
{\ddot t} + \frac{\tilde R}{R} {\dot t}^{2} = 0
\end{equation}

where the ~ denotes differentiation with respect to $t$.

We now treat the $k = -1 , 0, 1$ cases separately. In all the results
below we have chosen to use the $+$ sign in the R. H. S. of Eqn. (28).

(i) k = -1

This is the universe with a hyperbolic spacelike section. The scale factor
equation (28) has a solution given as :

\begin{equation}
R(t) = \frac{1}{C} \sinh (Ct + C_{1})
\end{equation}

where $C_{1}$ is an arbitrary integration constant. Thus the universe
begins with a big--bang (at $ t= -\frac{C_{1}}{C}$ and expands ever after.

The $r(\tau)$ and $t(\tau)$ turn out to be :

\begin{equation}
r(\tau) =\left [ (\frac{ C^{2}}{C_{0}} \tau)^{2} - 1\right ]^{-\frac{1}{2}}
\end{equation}

\begin{equation}
t(\tau) = \frac{1}{C} \left (\cosh^{-1} \frac{C^{2}}{C_{0}}\tau - C_{1}\right )
\end{equation}

where $C_{0}$ is an arbitrary integration constant.
The closed null string collapses at $\tau \rightarrow \pm \infty$. However
it is necessary that $\tau > \frac{C_{0}}{C^{2}}$

(ii) k = 0

Here the spacelike sections are flat. Eqn. (28) has a
solution which represents the inflationary universe, with a scale
factor given as :

\begin{equation}
R(t) = \exp (Ct)
\end{equation}

The string configuration is given as :

\begin{equation}
t(\tau) = \frac{1}{C}\ln (C\tau + C_{1})
\end{equation}
\begin{equation}
r(\tau) = \frac{1}{C(C\tau + C_{1})}
\end{equation}

Thus as $\tau \rightarrow \infty , t\rightarrow \infty$ and $r\rightarrow
0$. In this case also the closed null string collapses but only
as $\tau \rightarrow \infty$.

(iii) k = 1

This represents an universe with $S^{3}$ spacelike sections.
The scale factor has a solution given as :

\begin{equation}
R(t) = \frac{1}{C} \cosh (Ct + C_{1})
\end{equation}

Note that the scale factor is such that the universe is nonsingular
(there is no big--bang here).

The closed null string solution turns out to be :

\begin{equation}
r(\tau) =\left [ (\frac{ C^{2}}{C_{0}} \tau )^{2} + 1\right ]^{-\frac{1}{2}}
\end{equation}

\begin{equation}
t(\tau) = \frac{1}{C} \left (- C_{1} + \sinh^{-1} \frac{C^{2}}{C_{0}}\tau
\right )
\end{equation}

In this case also the string collapses only as $\tau\rightarrow \pm \infty$.

It can be seen quite easily that equations similar to (28) and (29)
(with the assumption $r(\tau)R(t) = C$) hold for a more general class of
cosmological spacetime metrics
generically represented as :

\begin{equation}
ds^{2} = = - dt^{2} + R^{2}(t)\left ( \frac{dr^{2}}{1-\frac{b(r)}{r}}
+ r^{2}d\theta^{2} + r^{2}\sin^{2}{\theta} d\phi^{2} \right )
\end{equation}

The Eqn. (29) remains unaltered whereas (28) becomes:

\begin{equation}
\tilde R = \pm C R \sqrt {1-\frac{b(r)}{r}}
\end{equation}

Thus for a given $b(r)$ we can solve (40) to find $R(t)$ and thereby
determine
$t(\tau)$ and $r(\tau)$. We now illustrtate this with two representative
examples.

(i)\hspace{.2in} First, let us choose $b(r) = \frac{b_{0}^2}{r}$.
 This represents
an evolving version of Ellis geometry {\cite{ellis:jmp73}}. The scale factor
equation has a solution given by :

\begin{equation}
R(t) = \frac{4 e^{Ct + C_{1}}}{1+ 4\alpha^{2}e^{2Ct + 2C_{1}}}
\end{equation}

The string solution turn out to be :

\begin{equation}
t = \frac{1}{C} \left \{ \ln \frac{1}{2\alpha}{\tan
\frac{\alpha C\tau}
{2}} - C_1  \right \}
\end{equation}

\begin{equation}
r(\tau) = \frac{\alpha}{C\sin \alpha C\tau}
\end{equation}

In the above $\alpha^{2} = b_{0}^{2}C^{2} $.

This represents a string which is significantly large at small
$\tau$ ($\tau \rightarrow 0 $) but becomes smaller and smaller
as one approaches  $\tau = \frac{\pi}{2\alpha C}$ where
$r = b_{0}$ and $t = \frac{1}{C}\left (\ln \frac{1}{2\alpha} - C_{1}
\right )$, $R(t) = \frac{1}{\alpha}$. Since $r(\tau)$ and $t(\tau)$
are both periodic in $\tau$ the solution is valid only in the domain
$\frac{n\pi}{\alpha C}<\tau<\frac{(n+1)\pi}{\alpha C}$.

(ii)\hspace{.3in} This case involves the evolving version of the
horizonfree Schwarzschild wormhole
{\cite{mt:ajp88}} for which we have $b(r) = b_{0}$.

The scale factor equation has a solution given by :

\begin{equation}
R(t) = \frac{4e^{Ct+C_{1}}}{(1+ \beta e^{Ct+ C_{1}})^{2}}
\end{equation}

The string configuration is :

\begin{equation}
t = \frac{1}{C} \left [ \ln \left (\frac{1}{\beta}\left \{\frac{4}{\beta C
(C_{2} - \tau)} - 1 \right \} \right ) - C_{1} \right ]
\end{equation}

\begin{equation}
r(\tau) = \frac{\beta}{4} \frac{(\frac{4}{\beta C(C_{2} -\tau)})^{2}}
{\frac{4}{\beta C(C_{2} -\tau)}- 1}
\end{equation}

where $\beta = b_{0} C$

This configuration is also defined only for $\tau < C_{2} -\frac{4}{\beta C}$
One can try out various choices for $b(r)$ and easily
obtain string configurations
and their corresponding scale factors by utilizing the relations described
in the previous paragraphs (Eqns (29) and (40)).

Alternatively, in Eqns (29) and (40) one can use $R(t)$ as the input
and derive the resulting functional form of $b(r)$(i.e. the features of
the spacelike hypersurface). For example, assuming $R(t) \sim t^{\nu}$
(which is motivated by the standard matter ($\nu = \frac{2}{3}$)
and radiation dominated ($\nu = \frac{1}{2}$)
scale factors) we obtain the following expressions for $t(\tau)$, $r(\tau)$
and $b(r)$.

\begin{equation}
t(\tau) = \left ( \frac{\nu + 1}{C_{0}}\right )^{\frac{1}{\nu +1}} (\tau -
C_{1})^{\frac{1}{\nu + 1}} \hspace{.2in} ; \hspace{.2in}
r(\tau) = \frac{1}{C} \left ( \frac{\nu + 1}{C_{0}}\right )^{\frac{-\nu}
{\nu +1}} (\tau - C_{1})^{\frac{-\nu}{\nu + 1}}
\end{equation}

and

\begin{equation}
b(r) = r\left ( 1 - {\nu}^{2}C^{\frac{2- 2\nu}{\nu}}r^{\frac{2}{\nu}}\right )
\end{equation}

For $\nu = \frac{1}{2}$ (radiation dominated FRW), $\nu = \frac{2}{3}$
(matter dominated FRW) and $\nu = 1$ (Milne) the $b(r)$ turns out to be

\begin{eqnarray}
b(r) = r \left ( 1 - \frac{1}{4} C^{2} r^{4}\right )\hspace{.3in} (\nu
= \frac{1}{2})\\
b(r) =  r \left ( 1 - \frac{4}{9}C^{2} r^{3}\right )\hspace{.3in} (\nu
= \frac{2}{3})\\
b(r) =  r \left ( 1 - r^{2}\right )\hspace{.3in} (\nu
= 1)
\end{eqnarray}

Unfortunately, none of these correspond to the standard or well known
spacelike hypersurfaces we encounter in the context of cosmology.

To conclude, let us now summarize the results obtained.

(1) We have written down the null string equations of motion and constraints
in general curved spacetimes. Specialising to specific backgrounds
we have also obtained explicit null string configurations.

(2) In spacetimes with event--horizons there is always a closed
, stationary null string on the horizon. In Rindler spacetime we have
been able to construct an explicit example of a dynamic string
configuration. It is possible that such configurations also
exist in general black hole spacetimes but we have'nt succeeded
in finding one.

(3) In cosmological backgrounds a general prescription has been
outlined following which one can construct a dynamic string configuration
in a fairly large class of such spacetimes. We have found string
configurations in the inflationary/de Sitter universes and some
other cases involving some interesting but not so popular
scale factors. Examples of string configurations in evolving
wormhole spacetimes have also been discussed towards the end.

It remains to be seen whether one can check the perturbative stability
of these string configurations. However, to do such an analysis we have
to set up a
formalism for null strings along the lines of the one for the timelike ones
(see  Guven,
{\cite{guv:prd93}}, Larsen and Frolov {\cite{lp:npb94}}, Carter
{\cite{cart:prd93}} and Capovilla and Guven{\cite{cap:prd95}}).

The existence of a closed stationary null string on the black hole
horizon may also have some nontrivial implications on the interpretation
of black--hole entropy from null string theory. Finally, it would
be worthwhile to work out in detail the conditions
under which general curved background is allowed in null string
theory. This is analogous to the $\beta$--function equations
for the timelike string. However, we have to be careful in the case of the
null string because we {\em cannot} use the inverse of the
 string tension as a
perturbation parameter here. On the contrary, since null strings
are tensionless and therefore correspond to the extreme high energy
limit of the tensionful theory one might be tempted to ask--
--does null string theory have a low--energy limit at all?
These and other issues are currently under investigation and will
be communicated in future.

It is a pleasure to thank Jnanadeva Maharana for making me aware
of several important references on null strings and  for
useful discussions. Financial support from the Institute
of Physics, Bhubaneswar is also gratefully acknowledged.

\end{document}